\documentclass[12pt]{article}

\usepackage[margin=1in]{geometry}
\usepackage{amsmath,amssymb,amsfonts}
\usepackage{hyperref}

\usepackage{times}
\usepackage[T1]{fontenc}
\usepackage{graphics, graphicx}
\usepackage{color}
\usepackage{subfigure}
\usepackage{textpos}
\usepackage[english]{babel}
\usepackage{url}

\DeclareGraphicsExtensions{.pdf,.png,.jpg}

\usepackage{setspace}
\allowdisplaybreaks

\def\BibTeX{{\rm B\kern-.05em{\sc i\kern-.025em b}\kern-.08em
		T\kern-.1667em\lower.7ex\hbox{E}\kern-.125emX}}

\title{A Bayesian spatio-temporal nowcasting model for public health decision-making and surveillance}

\date{}

\author{David Kline, Ayaz Hyder, Enhao Liu,\\ Michael Rayo, Samuel Malloy, Elisabeth Root}

\begin{document}

\maketitle

\begin{abstract}
As COVID-19 spread through the United States in 2020, states began to set up alert systems to inform policy decisions and serve as risk communication tools for the general public. Many of these systems, like in Ohio, included indicators based on an assessment of trends in reported cases.  However, when cases are indexed by date of disease onset, reporting delays complicate the interpretation of trends.  Despite a foundation of statistical literature to address this problem, these methods have not been widely applied in practice.  In this paper, we develop a Bayesian spatio-temporal nowcasting model for assessing trends in county-level COVID-19 cases in Ohio.  We compare the performance of our model to the current approach used in Ohio and the approach that was recommended by the Centers for Disease Control and Prevention.  We demonstrate gains in performance while still retaining interpretability using our model.  In addition, we are able to fully account for uncertainty in both the time series of cases and in the reporting process.  While we cannot eliminate all of the uncertainty in public health surveillance and subsequent decision-making, we must use approaches that embrace these challenges and deliver more accurate and honest assessments to policymakers.

  
  \textbf{Abbreviations:} COVID-19 - Coronavirus Disease 2019, OPHAS - Ohio Public Health Alert System
  \end{abstract}

\newpage

The first cases of SARS-CoV-2 in the United States were reported in early March \cite{Definition}, though recent phylogenetic evidence suggests the first introductions occurred in January 2020 \cite{Wu2020b,Worobey2020,Fauver2020}. As COVID-19 spread throughout the country, states began to set up risk alert systems to support data-driven decision-making, improve government accountability, and communicate health risks to the public \cite{Resolve}. The goal of such systems is to provide clear and consistent messaging around the current state of the COVID-19 pandemic and help people adopt protective behaviors while policy-makers implement appropriate structural changes to mitigate spread. Risk or public health alert systems typically develop a series of indicators which use various sources of surveillance data \cite{Resolve}.  In some states, these systems were linked to specific policy actions \cite{Utah} while in others they serve more as a risk communication tool to inform local health departments and the general public \cite{OPHAS}.  In most systems, several key indicators are tied to the reporting of confirmed COVID-19 cases and their onset date of illness (i.e., the date an individual first began to have symptoms) \cite{CDC}. However, chronic delays in outbreak investigation and case reporting have led to challenges in estimating case-based indicators and communicating the situation in a location in near real-time.

Issues related to reporting lag or reporting delay are not a new challenge in public health surveillance \cite{Brookmeyer1989,Kalbfleisch1989,Lawless1994}.  It is quite common for reporting in infectious disease and vital statistics systems to not occur instantaneously with the onset or occurrence of the event of interest.  For infectious diseases, this delay can be due to: 1) a prolonged interval between the time an individual recognizes symptoms and is able to seek care and receive confirmatory testing, 2) administrative backlogs and delays in the acquisition, processing, and ultimate reporting of information, and 3) the length of time necessary to conduct a full case investigation.  However, particularly when facing a fast-moving epidemic, important decisions need to be made in real-time despite the fact that the most recent information is likely incomplete. This added uncertainty can reduce the confidence of both policymakers and the public in the public health decision-making process.  Methodology is needed to help provide a clearer picture to decision-makers in the face of the uncertainty from delays in reporting.

To address this issue and build on the foundational methodology \cite{Brookmeyer1989,Kalbfleisch1989,Lawless1994}, a relatively recent literature around ``nowcasting" has emerged for delayed reporting.  In contrast to forecasting which focuses on estimating what could happen in the future, nowcasting focuses on estimating what has already happened but has not yet been reported.  Nowcasting leverages historical patterns in reporting and trajectories of the disease outcome to estimate current counts given partially reported values.  To enhance model flexibility and interpretability, recent work \cite{Stoner2019,vandeKassteele2019,McGough2020} has extended prior work for nowcasting time series \cite{Hohle2014} and aberration detection \cite{Salmon2015} within a Bayesian framework. This work has been applied to estimate COVID-19 deaths in regions of the United Kingdom \cite{Seaman2020} and to incorporate spatial dependence \cite{Stoner2019b,Rotejanaprasert2020}. In addition, simulation modeling approaches have also been used for nowcasting \cite{Shaman2013,Santillana2015,Wu2020}. In contrast to much of the current epidemiological work that relies on the specification of splines to capture trends, Bayesian structural time series can be specified as hierarchical autoregressive processes \cite{Scott2014}. Given the link between autoregressive processes and infectious disease dynamics, we propose a spatial extension of the Bayesian structural time series model to nowcast county-level counts of confirmed COVID-19 cases in Ohio while accounting for reporting delay.

Despite prior and current literature on methods for accounting for reporting delay, these methods have not been fully embraced in practice.  The purpose of this paper is to highlight the critical need to account for reporting lag and other potential daily reporting patterns when assessing whether case rates are increasing. This is important because an increase in case rates is an indicator in many states' alert systems, including the Ohio Public Health Alert System (OPHAS) \cite{OPHAS}, and can also serve as an early warning signal of disease spread. We apply our method to OPHAS Indicator 2 which measures "an increasing trend of at least 5 consecutive days in overall cases by onset date over the last 3 weeks" \cite{OPHAS}. Ohio adopted a 21 day "look-back" period in an attempt to manually curtail the effect of reporting delays. We develop an extension of a Bayesian structural time series model that incorporates spatial dependence across counties and flexibly captures temporal dynamics with an autoregressive structure.  We use case data from earlier in the pandemic that is now fully reported so the true trends can be determined for each county in Ohio.  We then compare indicators based on the method currently used in Ohio, the method suggested by the Centers for Disease Control and Prevention (CDC) \cite{CDC}, and our Bayesian approach.  

\section*{METHODS}

\subsection*{Data}

We used data on confirmed cases of COVID-19 in the state of Ohio which are captured by the Ohio Disease Reporting System and reported publicly \cite{Dashboard}.  In Ohio, case investigation is done by local, typically county, health departments and entered into the state system. Confirmed cases are defined as individuals who have a positive result on a laboratory molecular amplification test \cite{Definition} or other approved testing methods. For each individual case, the system records the county of residence and the onset date of illness as determined by case investigators.  If onset date is unknown, the system records the earliest date associated with the record.  Onset date currently provides the index date for all reporting and analysis at the state-level in Ohio.  The reporting date is defined as the first date at which a case appears in the system and is often several days or possibly weeks after the onset date.  Thus, when examining case counts by onset date, counts for the most recent days are incomplete because of the delay between onset date and reporting date. The reporting delay can also be impacted by system strains due to case volume and daily variation in reporting that differ by local health department.

To explore the impact of reporting patterns on the calculation and subsequent interpretation of public health alert indicators, we retrospectively consider four points in time during the pandemic: June 15, 2020, July 15, 2020, August 15, 2020, and September 15, 2020.  At the time of the analysis, at least one full month had passed since September 15, and we assume that case reporting was complete through this date.  For each date, we examine cases reported by that date and compute indicators related to the trends in case counts. Since the data are completely reported, we can compare the estimates from the indicators to the true trend observed in the onset cases at that point in time.  This will allow us to examine the performance of each proposed approach for determining if a county is experiencing an increasing trend of cases.    

\subsection*{Rolling Average Approach}

We refer to the current approach for determining if case rates are increasing used by the Ohio Department of Health as the rolling average approach \cite{OPHAS}. This approach computes a 7 day rolling average of case counts, indexed by onset date, for each of the last 21 days.  The alert indicator for an increasing trend in cases is flagged if there are 5 consecutive days of increasing averages at any point in the 21 day window.  That is, the indicator flags if for 5 consecutive days the average is greater than the average the day before. This approach crudely accounts for daily reporting variation by averaging across 7 days but makes no attempt to account for reporting lag or any other sources of variation.

\subsection*{Spline Approach}

A slightly more sophisticated but still simple approach was recommended by the CDC for detecting rebounds \cite{CDC} and will be referred to as the spline approach.  This approach is similar to the rolling average approach described above but fits a spline to the time series of rolling averages.  For consistency, we used 7 day rolling averages over a 21 day period to align with the temporal window of interest for the alert system.  We fit a cubic spline \cite{CDC} to each series with 4 knots.  By using a spline, we are able to smooth daily and other systematic variation in reporting patterns.  Aligned with the CDC \cite{CDC}, we determine if there is an increasing trend by looking at the fitted values from the spline and determining if there are any 5 consecutive day periods where the fit for each day is greater than the previous day.  Like the rolling average approach, uncertainty is not incorporated into the decision-making process. Splines were estimated using the \texttt{mgcv} package in R \cite{Wood2017}.

\subsection*{Model-based Approach}
In contrast to the simpler approaches, we explicitly model both the process for new onset cases and the reporting delay process. We extend the general framework outlined by previous work \cite{Stoner2019,Stoner2019b} by using an autoregressive spatial Bayesian structural time series, rather than a spline based model.  While the spline based model is flexible, it relies on reasonably specifying knots and is not ideal for estimating beyond the range of the observed data.  In addition, it can be more challenging to incorporate hierarchical structure when temporal trends may be quite different across locations, which has been the case for COVID-19. Instead, an autoregressive structure retains the ability to flexibly capture spatio-temporal trends while also linking more closely to the dynamics of infectious disease \cite{Viboud2006}. It also allows for added flexibility in specifying a spatially varying reporting delay process.  

We follow the general set up outlined in previous work \cite{Stoner2019,Salmon2015}.  In Ohio, COVID-19 cases are reported daily so we use a daily time scale.  To reduce computation time, we will take a moving window approach \cite{McGough2020} that considers the past 90 days ($T=90$). From April through September 2020, 94\% of cases were reported within 2 weeks of onset and 98\% of cases were reported within 30 days.  To be conservative, we set a maximum reporting delay time of 30 days following onset ($D=30$).  

\textit{Outcome Model.} Let $Y_{it}$ be the count of reported cases in county $i=1,\ldots,N$ with onset date $t=1,\ldots,T$. Note that $Y_{it}$ is assumed to be the true total count, which is assumed to be partially observed for time $t$ such that $t+D>T$. We assume
\begin{align*}
    &Y_{it} \sim \text{Poisson}(\lambda_{it})\\
    &\text{log}(\lambda_{it})=O_i + \alpha_{it} + \textbf{X}_t \boldsymbol{\eta}_i
\end{align*}
where $O_i$ is an offset of the log population of county $i$, $\alpha_{it}$ is the latent state of the process, $\textbf{X}_t$ is a design vector indicating the day of the week, and $\boldsymbol{\eta}_i$ is the day of the week effect. Note that $\textbf{X}_t$ is parameterized using sum to 0 effect coding so $\alpha_{it}$ reflects the average of the process across days of the week.  By using this structure for the model, we are able to remove daily reporting variation from the latent state, $\alpha_{it}$, through $\textbf{X}_t \boldsymbol{\eta}_i$.

After removing the daily ``seasonal" variation, we focus on the model for the latent state or structural part of the model.  We use a semi-local linear trend model \cite{Brodersen2015} to allow for some degree of longer term structure while still facilitating a very flexible model.  That is for $t>1$,
\begin{align*}
    \alpha_{it} = \alpha_{i(t-1)}+\delta_{i(t-1)}+\epsilon^{\alpha}_{it}
\end{align*}
where $\epsilon^{\alpha}_{it} \stackrel{iid}{\sim} N(0,\tau^2_{\alpha})$ and the initial value at $t=1$ is $\alpha_{i1} \sim N(0,100)$.  Then for the model for trend, we let
\begin{align*}
    &\delta_{i1} = \delta + d_i +\epsilon^{\delta}_{i1}\\
    &\delta_{it} = \delta + d_i + \rho_{\delta} (\delta_{i(t-1)} - \delta - d_i) +\epsilon^{\delta}_{it}
\end{align*}
where $\delta$ is a common statewide trend, $d_i$ is a county-specific spatial trend, and $\rho_{\delta}$ is an autoregressive term.  Let $\epsilon^{\delta}_{it} \stackrel{iid}{\sim} N(0,\tau^2_{\delta})$.  A benefit to this parameterization is it allows us to separate changes that are due to white noise ($\epsilon^{\alpha}_{it}$) from those that are due to more consistent temporal trends ($\delta_{it}$). By using a stationary model for $\delta$, we are able to provide some structure around a longer term trend while retaining flexibility for local deviations in space and time.

To account for spatial correlation, we assume the trends in neighboring counties are correlated and specify an intrinsic conditional autoregressive model.  That is,
\begin{align*}
    d_i|d_{-i} \sim N\left(\frac{1}{w_{i+}} \sum_{j} w_{ij} d_{j}, \frac{\tau_d^2}{w_{i+}} \right)
\end{align*}
where $d_{-i}$ is the set of counties excluding county $i$, $w_{ij}$ is an indicator of whether counties $i$ and $j$ are adjacent, $w_{i+}=\sum_{j \neq i} w_{ij}$, and $\tau^2_d$ is a variance.  To ensure a valid process model, we enforce a sum to 0 constraint on the $d_i$ \cite{Banerjee2004}. We chose to incorporate spatial dependence in the trend to reflect a belief that cases in a county are likely to change in a similar fashion as cases in neighboring counties.  This choice explicitly aligns with our general surveillance and risk evaluation strategy for counties where we have implicitly considered trends in neighboring counties when making our assessments.  Another added benefit is that this helps to stabilize estimates for counties with small populations by borrowing strength from neighboring counties.

We also assume county-specific effects of the day of the week.  We assume that while variability exists between counties, the daily patterns are similar across the state.  We assume the following hierarchical model
\begin{align*}
    \boldsymbol{\eta}_i \stackrel{iid}{\sim} N(\boldsymbol{\eta},\tau^2_{\eta} \textbf{I}_6)
\end{align*}
where $\boldsymbol{\eta}_i$ is a vector of state average day of the week effects, $\tau^2_{\eta}$ is a variance, and $\textbf{I}_6$ is a $6 \times 6$ identity matrix. This allows each county to have its own daily pattern while borrowing strength across all counties in the state as warranted.

\textit{Reporting Model.} Since we know that $Y_{it}$ is observed with reporting lag, we must specify a model for the delay.  Let $Z_{itd}$ be the count of cases observed in county $i$ with onset date $t$ that are observed $d=0,\dots,D$ days after $t$. Note that $Z_{itd}$ corresponds to when cases are reported $d$ days after onset date $t$ and so is unobserved when $t+d>T$. We assume
\begin{align*}
    &\textbf{Z}_{it}|\textbf{p}_{it},Y_{it} \sim \text{Multinomial}(\textbf{p}_{it},Y_{it})\\
    &\textbf{p}_{it} \sim GD(\boldsymbol{\alpha}_{it},\boldsymbol{\beta}_{it})
\end{align*}
where $\textbf{Z}_{it}=(Z_{it0},\ldots,Z_{itD})$, $\textbf{p}_{it}$ is the vector of proportions of the total $Y_{it}$ reported on each of the $D$ days, and GD is the generalized Dirichlet distribution. We use a generalized Dirichlet distribution to properly account for potential overdispersion of the $\textbf{p}_{it}$ \cite{Stoner2019}. This leads to the following conditional distribution:
\begin{align*}
    Z_{itd}|Z_{it(-d)},Y_{it} \sim \text{Beta-Binomial}\left(\alpha_{itd},\beta_{itd},Y_{it}-\sum_{j<d}Z_{itj}\right)
\end{align*}
where $Z_{it(-d)}$ is set of counts reported with a delay that is not $d$ days.  To model more intuitive quantities, we reparameterize the distribution  \cite{Stoner2019} in terms of the mean $\nu_{itd}$ and dispersion $\phi_{d}$ such that $\alpha_{itd}=\nu_{itd} \phi_{d}$ and $\beta_{itd}=(1-\nu_{itd})\phi_{d}$.  Then similar to a hazard function, we let $\text{logit}(\nu_{itd})=\psi_{itd}$ and assume the following AR1 model
\begin{align*}
    &\psi_{i1d}=\beta_d + \textbf{V}_{1d} \boldsymbol{\xi}_i + \epsilon^{\psi}_{i1d}\\
    &\psi_{itd}=\beta_d + \textbf{V}_{td} \boldsymbol{\xi}_i + \rho_{\nu} (\psi_{i(t-1)d} - \beta_d + \textbf{V}_{(t-1)d} \boldsymbol{\xi}_i) + \epsilon^{\psi}_{itd}
\end{align*}
where $\beta_d$ is the average log odds of remaining cases being reported by delay $d$, $\textbf{V}_{td}$ is a design matrix indicating the day of the week, $\boldsymbol{\xi}_i$ is a day of the week effect, $\rho_{\psi}$ is an autoregressive parameter, and $\epsilon^{\psi}_{itd}$ is an error term.  We assume $\epsilon^{\psi}_{itd} \stackrel{iid}{\sim} N(0,\tau^2_{\psi})$. Note that $\textbf{V}_{td}$ is parameterized using sum to 0 effect coding.

The parameterizaton of the delay model allows us to accommodate several important features of COVID-19 reporting and should, in general, be customized to reflect the actual reporting process.  First, reporting in Ohio is done by county health departments who may have varying capacity and resources for timely reporting.  Thus, the delay model is county-specific.  We account for day of the week effects, much like in the model for the case counts, because in many counties, reporting primarily aligns with the work week.  We also assume autoregressive temporal dependence to capture the potential for administrative backlogs.  For example, if a smaller portion of cases are reported today, we may also expect a smaller proportion the next day because of a backlog.  We do not incorporate a term to account for spatial dependence in the delay model as we assume neighboring health departments are independent agencies, and so we would not anticipate spatial structure.

As with the outcome model, we allow for county-specifc variability in day of the week reporting effects.  We again assume similar patterns across the state and specify the following hierarchical model:
\begin{align*}
    \boldsymbol{\xi}_i \stackrel{iid}{\sim} N(\boldsymbol{\xi},\tau^2_{\xi} \textbf{I}_6)
\end{align*}
where $\boldsymbol{\xi}$ is a vector of state average day of the week effects, $\tau^2_{\xi}$ is a variance, and $\textbf{I}_6$ is a $6 \times 6$ identity matrix.

\textit{Prior Model and Computation.} Since we fit our model in the Bayesian paradigm, we must specify prior distributions on all unknown parameters. For each element of $\boldsymbol{\eta}$ and $\boldsymbol{\xi}$, we assign independent normal priors with 0 mean and variance 1.  We also assign $\delta$ a normal prior with 0 mean and variance 1.  We use a variance of 1 for these prior distributions as each parameter reflects a relative daily difference on the log scale, and so these priors reflect a reasonable range for those parameters.  We assign $\beta_d$ independent normal priors with mean 0 and variance 4, which puts adequate probability on reasonable values on the logit scale. We also assign all variance parameters inverse gamma priors with shape and scale both set to 0.5.  All autoregressive parameters are assigned uniform prior distributions over -1 to 1.

To compare across approaches, we fit the model for each of the four dates considered. We treat the last day in the series (i.e., the current date) as missing and forecast the expected case count, which reduces model instability due to the rarity of cases reported on the day of onset ($d=0$). The model was fit using a Markov chain Monte Carlo algorithm implemented in R using \texttt{nimble} \cite{nimble}.  The algorithm was run for 30,000 iterations with the first 15,000 discarded as burn-in and then thinned by keeping every 10th iteration. Computation time was approximately 20 hours, which would enable a daily update in practice.

To determine whether the cases were increasing in the most recent 21 day period, we use the posterior distribution of $\delta_{it}$.  Since $\delta_{it}$ reflects the trend in county $i$ at time $t$, there is a net increasing trend over the past 21 days if $\sum_{t=T-20}^{T} \delta_{it}>0$.  Using the posterior distribution, we can directly compute the posterior probability of an increasing trend for each county.

\subsection*{True Change}

One major advantage of a model-based approach is the flexibility to address more complex questions of interest.  However, the goal of this paper is to assess the method used to calculate the OPHAS indicator for when cases are increasing in a county.  To most closely align with the question as currently posed by the state of Ohio, we define a true increase in cases as when the number of cases in the most recent 7 day period is greater than the number of cases two weeks prior.  This corresponds to comparing the first week with the last week in the most recent 21 day period.  While there are other potential ways to define a true increase, this most closely reflects the current definition used by the state of Ohio.

\section*{RESULTS}

\begin{figure}
	\centering
	\subfigure[June 15, 2020]{
		\label{fig:comp1}
		\includegraphics[width=\textwidth]{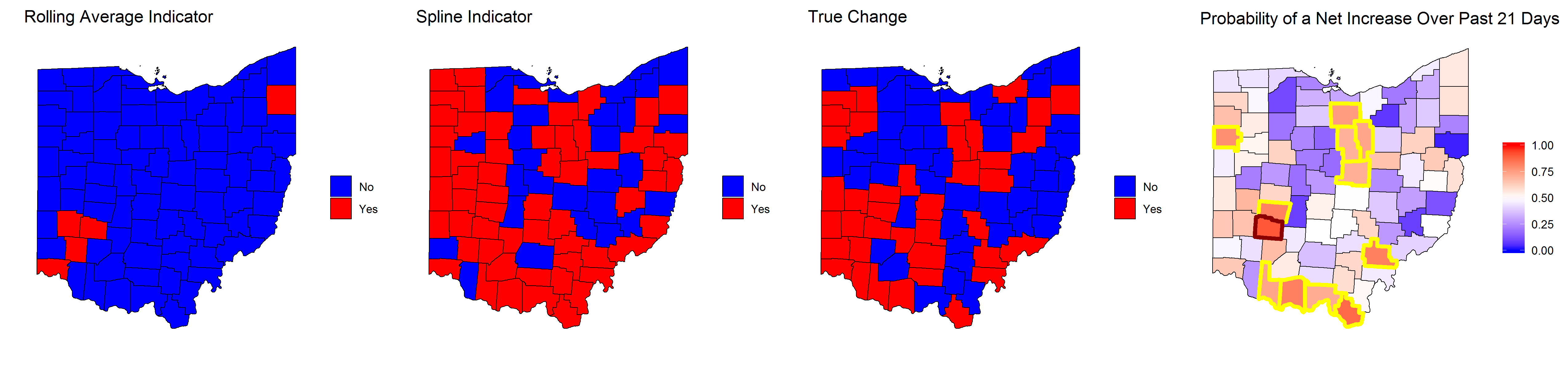}
	}
	\subfigure[July 15, 2020]{
		\label{fig:comp2}
		\includegraphics[width=\textwidth]{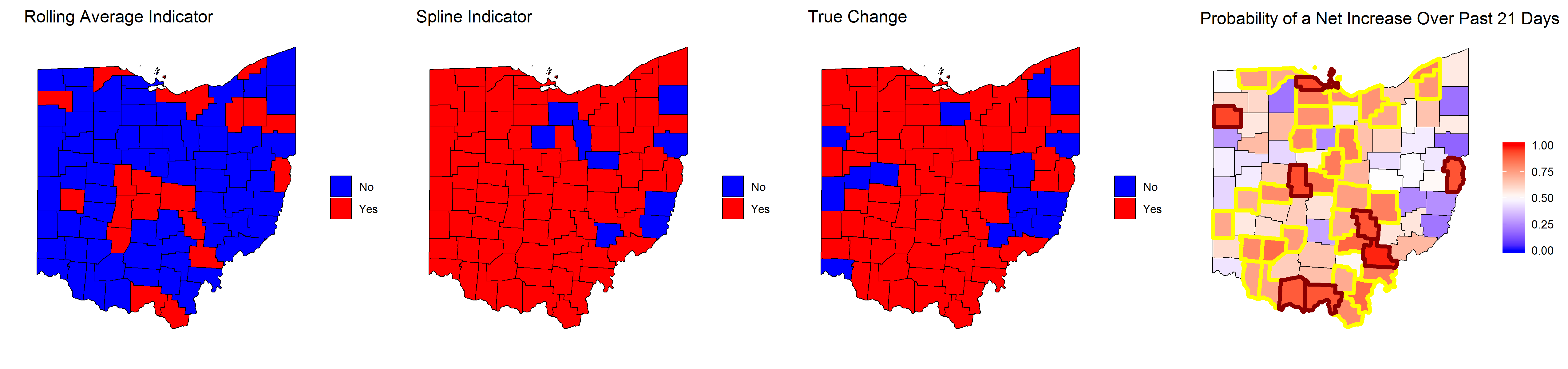}
	}
    \subfigure[August 15, 2020]{
		\label{fig:comp3}
		\includegraphics[width=\textwidth]{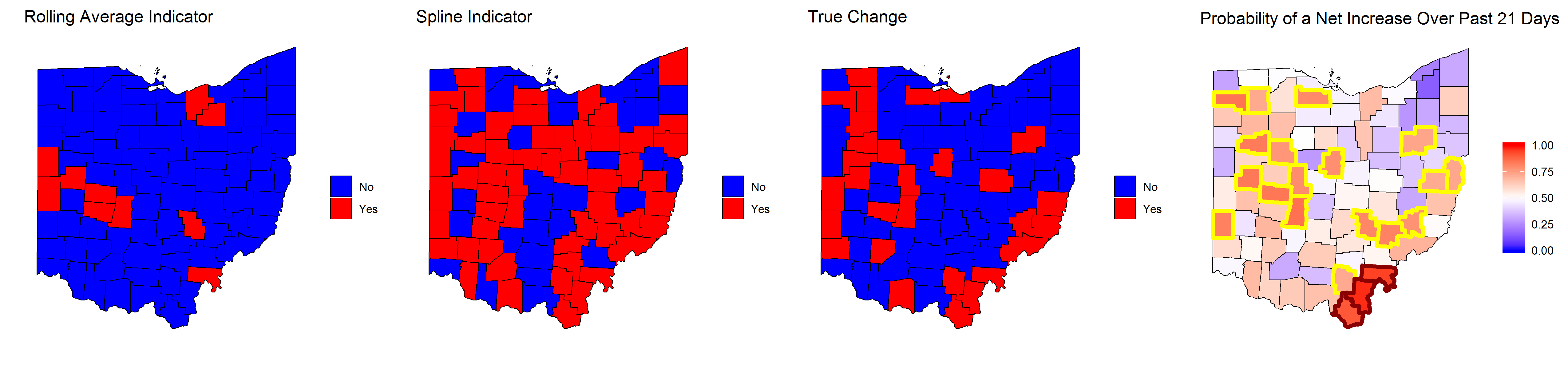}
	}
    \subfigure[September 15, 2020]{
		\label{fig:comp4}
		\includegraphics[width=\textwidth]{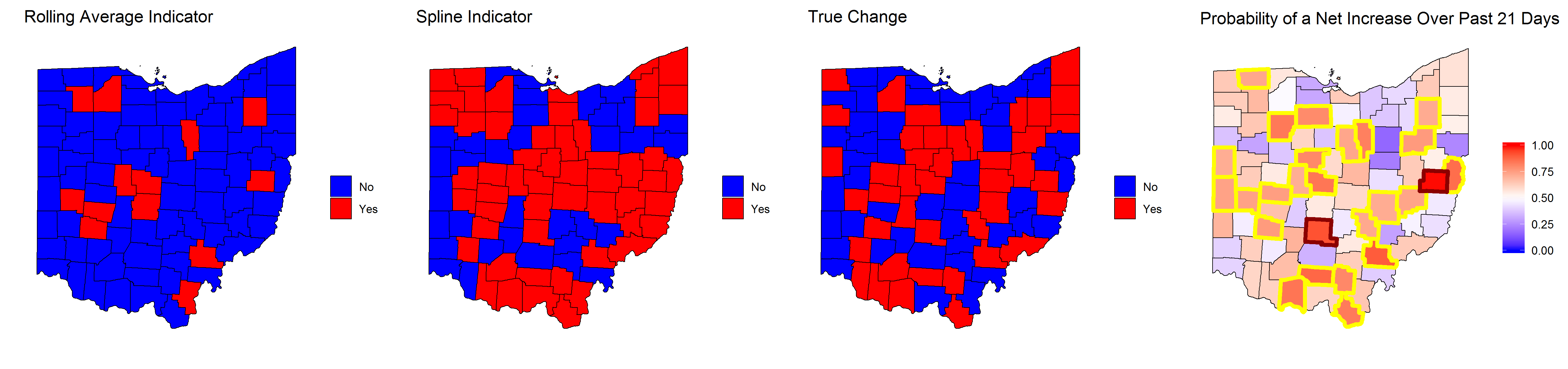}
	}
	\caption[Comparisons]{Comparison between the rolling average indicator, the spline indicator, the true observed indicator of an increase, and the model-based posterior probability across 4 time points during the pandemic. For the model-based probabilities, counties outlined in red have a probability greater than 0.9 and outlined in yellow have a probability greater than 0.7.}
	\label{fig:est}
\end{figure}

The results from applying each of the three methods for calculating increasing case rates are shown in Figure \ref{fig:est}.  There are several general observations that can be made across the four time points. The rolling average indicator generally does a poor job at accurately capturing counties where the cases have increased, and in most counties, there were true increases that went undetected.  The spline indicator tends to make errors in the other direction by incorrectly flagging counties that did not meet the definition of a true increase. For the model-based approach, we generate a posterior probability of an increasing trend and highlight counties in yellow with a probability greater than 0.7 and in red those with a probability greater than 0.9.  

\begin{table}[t]
	\caption{\label{table:sens_spec} Estimated sensitivity and specificity of the rolling average indicator, the spline indicator, and the model-based indicator at 3 different posterior probability cut-points across the 4 dates examined.}
    \centering
	\begin{tabular}{lcc}
    	\hline
		Method & Sensitivity & Specificity\\
		\hline
		Rolling Average & 0.20 & 0.96 \\
		Spline & 0.87 & 0.48 \\
		Model-based: >0.9 & 0.07 & 1.00 \\
		Model-based: >0.7 & 0.46 & 0.93 \\
		Model-based: >0.5 & 0.83 & 0.60 \\
        \hline
	\end{tabular}
\end{table} 

In addition to visually examining the results, we calculated sensitivity and specificity for each approach in Table \ref{table:sens_spec}.  The rolling average approach currently in use has a very low sensitivity of 0.20 and so is not successfully identifying counties with increasing trends.  The spline approach has a much higher sensitivity of 0.87 but at the cost of a specificity of 0.48.  Three cut points are shown for the model-based posterior probabilities.  As expected, the higher thresholds exhibit excellent specificity but lower sensitivity since it reflects stronger evidence of an increase.  Using a cut-point of 0.5, which reflects that the trend is more likely increasing than decreasing, we estimate a sensitivity of 0.83 and a specificity of 0.6, which seems to most reasonably balance false positives and false negatives among the approaches considered.  

\begin{figure}
	\centering
    \subfigure[Franklin County - urban]{
		\label{fig:urban}
		\includegraphics[width=\textwidth]{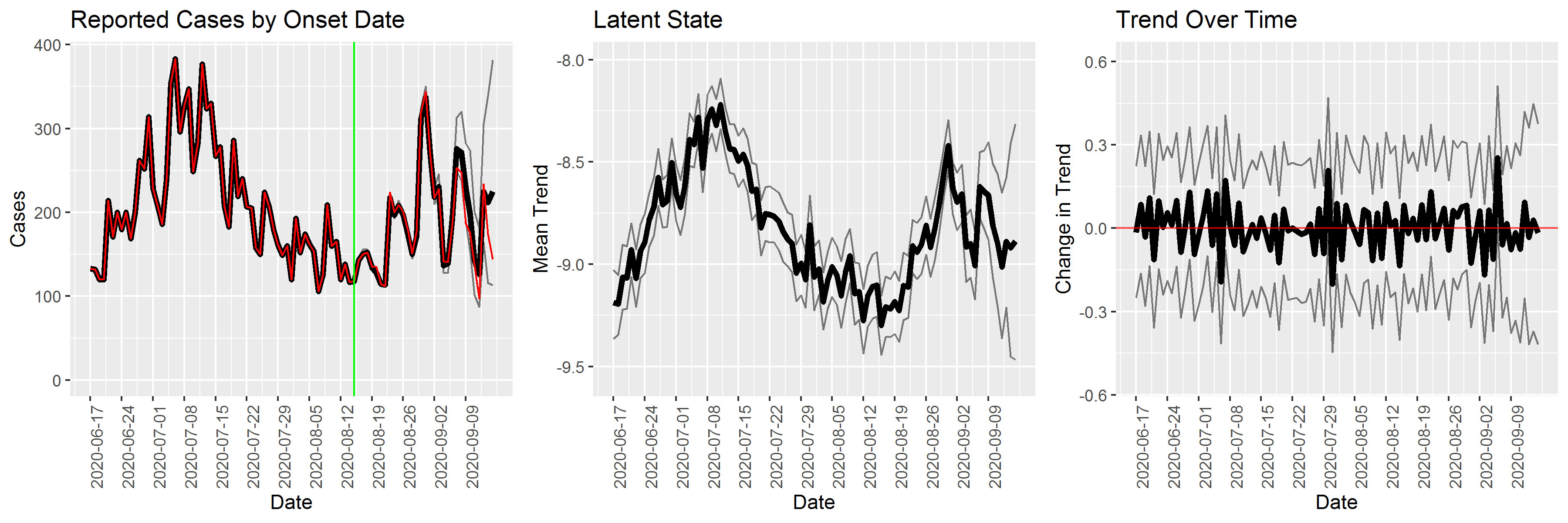}
	}
    \subfigure[Harrison County - rural]{
		\label{fig:rural}
		\includegraphics[width=\textwidth]{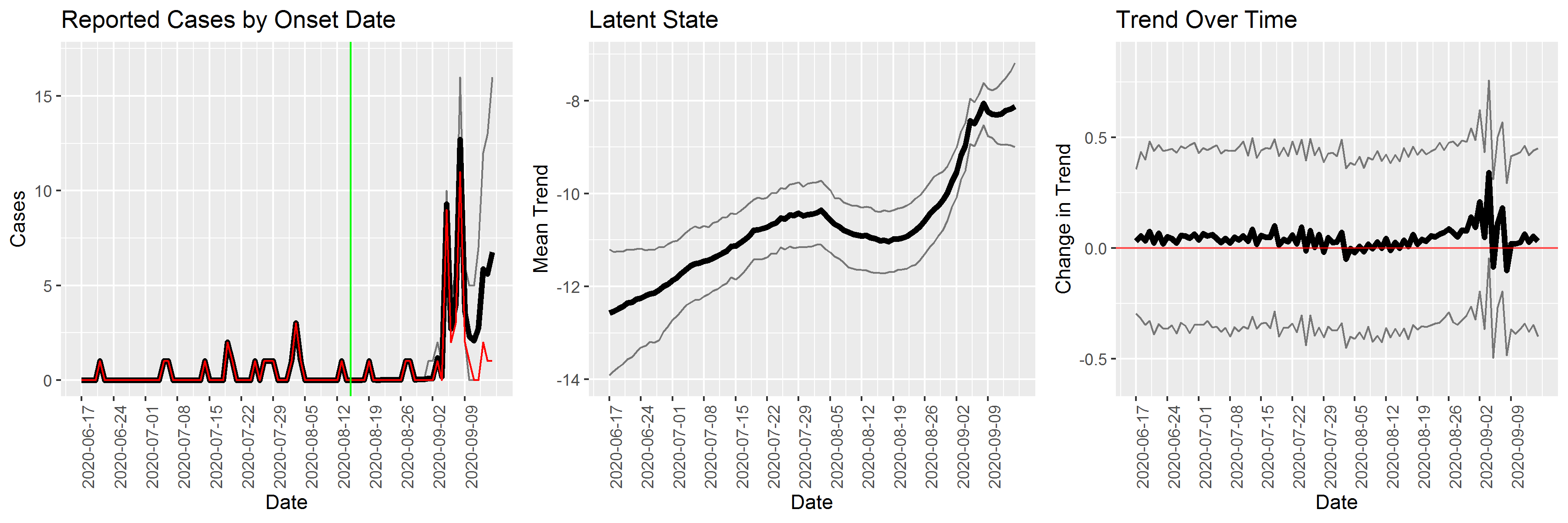}
	}
	\caption[Nowcast]{Case nowcast projections and time series model components for an urban and rural county on September 15, 2020. The left panel shows the posterior mean number of cases in bold, 90\% credible interval in black, and the true number of cases in red.  The green vertical line indicates the divide between complete and incomplete reporting. The center panel shows the posterior mean and 90\% credible interval of the latent state which is the mean process on the log scale with daily variation removed. The right panel shows the posterior mean and 90\% credible interval for the daily change with a red reference line at 0.}
	\label{fig:case}
\end{figure}

The model-based approach also provides a rich set of additional results that can provide useful insights.  Typically, the main goal of these models is to nowcast case counts. In Figure \ref{fig:case}, we show nowcast estimates with their 90\% credible interval in black and the true counts in red for an urban and rural county.  Averaging across the 4 time points, the 90\% credible interval coverage was 0.96 over the 30 day period with incomplete reporting. The coverage was 0.92 in the most recent 7 days which have the most incomplete reporting.  Thus, our model performs as expected for nowcasting cases. In Figure \ref{fig:case}, we also show time series plots of the latent state, which removes the daily seasonality, and the trend.  The trend can also be viewed as the derivative of the latent state curve so when it is greater than 0, it indicates increasing case counts. 

\section*{DISCUSSION}

We applied three approaches for assessing increasing trends in cases to completely observed data at four time points during the COVID-19 pandemic.  When assessments are linked to onset date, case reporting is subject to reporting lag or delay.  We illustrate that the simple approach currently used in OPHAS does not perform well as it fails to account for lag and other variation in reporting.  The spline approach outlined by the CDC is more sensitive as it smooths over daily reporting variation but also fails to account for lag.  In contrast, the model-based approach accommodates lag, daily variation, and spatio-temporal dependence.  The model-based approach can also directly summarize observed evidence of increasing trends and the associated uncertainty through posterior distributions.  This results in a better trade-off between sensitivity and specificity and can allow for prioritization of areas where the evidence of an increase is strongest.

We note several key advantages to the model-based approach. First, the Bayesian approach allows us to use calculated posterior summaries to directly communicate uncertainty. Public health officials are constantly considering trade offs between different policy options - e.g., stay-at-home orders vs. economic impacts. Specific policy responses may only be warranted when evidence for an increase in COVID-19 cases is very strong and models indicate a very high level of certainty. Since the posterior probability reflects the probability of an increasing trend given the observed data, this quantity can be used to directly address the policy question of interest and provides an indication of how strong the evidence is in each county. Unlike the spline or 7-day rolling average approaches that return a binary decision, the ability of the model-based approach to convey additional meaning through continuous estimates is a clear advantage that can improve decision-making \cite{Rayo2015a,Rayo2015b}. Second, by accounting for reporting delays and fully exploiting partially reported counts, the Bayesian approach can be more responsive to changing trends and provide earlier warning of changes in trends. Finally, the output from the Bayesian models (shown in Figure \ref{fig:case}) provide important additional information that can be used by surveillance teams to understand trends over time. These results do require a team of epidemiologists to review the data, but still provide more information than the spline or 7-day moving average methods.

When responding to a pandemic, it is important that the public health and policy response is guided by the best available information.  Often even the best information can be incomplete and uncertain.  However, statistical models have been developed to overcome these issues and aid in characterizing and quantifying uncertainty.  These models are not as simple as the approach currently used in Ohio, and this is one limitation of this method. Risk alert systems should be transparent and easy to understand. Complex modeling approaches are difficult to explain to the general public and can lead to mistrust in the data and, by extension, the system as a whole. However, with proper preparation, the model output can be summarized to simply communicate the core messages, while leaving much of the complexity and technical details to the experts implementing the model. Additionally, the Bayesian models provide a wide range of information that can be used internally by epidemiologists and other public health data scientists to directly address important policy questions. Given the clear improvements our Bayesian models offer, it is imperative that we take advantage of these methodological advances to better serve the public and inform the distribution of limited resources.


In conclusion, we have illustrated shortcomings in using simple approaches for public health decision-making. We have also illustrated how more sophisticated statistical models can account for the real-world complexities associated with surveillance data.  Despite the added complexity, the output from these models can be summarized in a relatively simple and concise form that still appropriately reflects uncertainty.  While we cannot eliminate all of the uncertainty in public health surveillance and decision-making, we must use approaches that embrace these challenges and deliver more accurate and honest assessments to policymakers.

\bibliography{lag_refs} 

\begin{thebibliography}{10}

\bibitem{Definition}
{Centers for Disease Control and Prevention} . Coronavirus Disease 2019
  ({COVID}-19) 2020 Interim Case Definition, Approved {A}ugust 5, 2020   2020.
\newblock
  \url{https://wwwn.cdc.gov/nndss/conditions/coronavirus-disease-2019-covid-19/case-definition/2020/08/05/}.

\bibitem{Wu2020b}
Wu~S.L., Mertens A.N., Crider Y.S., et al. Substantial underestimation of
  {SARS-CoV-2} infection in the {U}nited {S}tates  {\it Nature Communications.
  } 2020;11:1--10.

\bibitem{Worobey2020}
Worobey Michael, Pekar Jonathan, Larsen Brendan~B., et al. The emergence of
  SARS-CoV-2 in Europe and North America  {\it Science. } 2020;370:564--570.

\bibitem{Fauver2020}
Fauver Joseph~R., Petrone Mary~E., Hodcroft Emma~B., et al. Coast-to-Coast
  Spread of SARS-CoV-2 during the Early Epidemic in the United States  {\it
  Cell. } 2020;181:990 - 996.e5.

\bibitem{Resolve}
{Resolve to Save Lives} . Staying Alert: Navigating {COVID}-19 Risk Toward a
  New Normal   2020.
\newblock
  \url{https://preventepidemics.org/wp-content/uploads/2020/05/STAYING-ALERT-Navigating-COVID-19-Risk-Toward-a-New-Normal_final.pdf}.

\bibitem{Utah}
{Utah Department of Health} . Phased Guidelines for the General Public and
  Businesses to Maximize Public Health and Economic Reactivation   2020.
\newblock
  \url{https://coronavirus-download.utah.gov/Health/Phased\%20Health\%20Guidelines\%20V4.0.1.pdf}.

\bibitem{OPHAS}
{Ohio Department of Health} . Summary of Alert Indicators   2020.
\newblock
  \url{https://coronavirus.ohio.gov/static/OPHASM/Summary-Alert-Indicators.pdf}.

\bibitem{CDC}
{Centers for Disease Control and Prevention} . {CDC} Activities and Initiatives
  Supporting the {COVID}-19 Response and the President’s Plan for Opening
  America Up Again   2020.
\newblock
  \url{https://www.cdc.gov/coronavirus/2019-ncov/downloads/php/CDC-Activities-Initiatives-for-COVID-19-Response.pdf}.

\bibitem{Brookmeyer1989}
Brookmeyer R., Damiano A.. Statistical methods for short-term projections of
  AIDS incidence  {\it Statistics in Medicine. } 1989;8:23-34.

\bibitem{Kalbfleisch1989}
Kalbfleisch J., Lawless J.. Inference Based on Retrospective Ascertainment: An
  Analysis of the Data on Transfusion-Related AIDS  {\it Journal of the
  American Statistical Association. } 1989;84:360-372.

\bibitem{Lawless1994}
Lawless J.F.. Adjustments for reporting delays and the prediction of occurred
  but not reported events  {\it Canadian Journal of Statistics. }
  1994;22:15-31.

\bibitem{Stoner2019}
Stoner O., Economou T.. Multivariate hierarchical frameworks for modeling
  delayed reporting in count data  {\it Biometrics. } 2020;76:789-798.

\bibitem{vandeKassteele2019}
Kassteele J., Eilers P.H., Wallinga J.. Nowcasting the Number of New
  Symptomatic Cases During Infectious Disease Outbreaks Using Constrained
  P-spline Smoothing  {\it Epidemiology. } 2019;30:737--745.

\bibitem{McGough2020}
McGough S.F., Johansson M.A., Lipsitch M., Menzies N.A.. Nowcasting by Bayesian
  Smoothing: A flexible, generalizable model for real-time epidemic tracking
  {\it PLOS Computational Biology. } 2020;16:1-20.

\bibitem{Hohle2014}
Hohle M., Heiden M.. Bayesian nowcasting during the STEC O104:H4 outbreak in
  Germany, 2011  {\it Biometrics. } 2014;70:993-1002.

\bibitem{Salmon2015}
Salmon M., Schumacher D., Stark K., Höhle M.. Bayesian outbreak detection in
  the presence of reporting delays  {\it Biometrical Journal. }
  2015;57:1051-1067.

\bibitem{Seaman2020}
Seaman S., Samartsidis P., Kall M., De~Angelis D.. Nowcasting CoVID-19 Deaths
  in England by Age and Region  {\it medRxiv. } 2020.

\bibitem{Stoner2019b}
Stoner O., Economou T.. A Hierarchical Modelling Framework for Correcting
  Delayed Reporting in Spatio-Temporal Disease Surveillance Data  {\it arXiv. }
  2019.

\bibitem{Rotejanaprasert2020}
Rotejanaprasert C., Ekapirat N., Areechokchai D., Maude R.J.. Bayesian
  spatiotemporal modeling with sliding windows to correct reporting delays for
  real-time dengue surveillance in Thailand  {\it International Journal of
  Health Geographics. } 2020;19.

\bibitem{Shaman2013}
Shaman Jeffrey, Karspeck Alicia, Yang Wan, Tamerius James, Lipsitch Marc.
  Real-time influenza forecasts during the 2012–2013 season  {\it Nature
  Communications. } 2013;4.

\bibitem{Santillana2015}
Santillana Mauricio, Nguyen André~T., Dredze Mark, Paul Michael~J., Nsoesie
  Elaine~O., Brownstein John~S.. Combining Search, Social Media, and
  Traditional Data Sources to Improve Influenza Surveillance  {\it PLOS
  Computational Biology. } 2015;11:1-15.

\bibitem{Wu2020}
Wu~Joseph~T, Leung Kathy, Leung Gabriel~M. Nowcasting and forecasting the
  potential domestic and international spread of the 2019-nCoV outbreak
  originating in Wuhan, China: a modelling study  {\it The Lancet. }
  2020;395:689--697.

\bibitem{Scott2014}
Scott S., Varian H.. Predicting the present with Bayesian structural time
  series  {\it Int. J. Math. Model. Numer. Optimisation. } 2014;5:4-23.

\bibitem{Dashboard}
{Ohio Department of Health} . {COVID}-19 Dashoard   2020.
\newblock
  \url{https://coronavirus.ohio.gov/wps/portal/gov/covid-19/dashboards/overview}.

\bibitem{Wood2017}
Wood S.N. {\it Generalized Additive Models: An Introduction with R}.
\newblock Chapman and Hall/CRC2~ed. 2017.

\bibitem{Viboud2006}
Viboud C{\'e}cile, Bj{\o}rnstad Ottar~N., Smith David~L., Simonsen Lone, Miller
  Mark~A., Grenfell Bryan~T.. Synchrony, Waves, and Spatial Hierarchies in the
  Spread of Influenza  {\it Science. } 2006;312:447--451.

\bibitem{Brodersen2015}
Brodersen K.H., Gallusser F., Koehler J., Remy N., Scott S.L.. Inferring causal
  impact using Bayesian structural time-series models  {\it Ann. Appl. Stat.. }
  2015;9:247--274.

\bibitem{Banerjee2004}
Banerjee Sudipto., Carlin Bradley~P., Gelfand Alan~E.. {\it Hierarchical
  modeling and analysis for spatial data}.
\newblock Boca Raton, Fla.: Chapman \& Hall/CRC 2004.

\bibitem{nimble}
{de Valpine} P., Turek D., Paciorek C.J., Anderson-Bergman C., {Temple Lang}
  D., Bodik R.. Programming with models: writing statistical algorithms for
  general model structures with {NIMBLE}  {\it Journal of Computational and
  Graphical Statistics. } 2017;26:403-417.

\bibitem{Rayo2015a}
MF~Rayo, N~Kowalczyk, BW~Liston, S~White, ES~Patterson. Comparing the
  Effectiveness of Alerts and Dynamically Annotated Visualizations (DAVs) in
  Improving Clinical Decision Making  {\it Human Factors: the Journal of the
  Human Factors and Ergonomics Society. } 2015;57:1002--1014.

\bibitem{Rayo2015b}
MF~Rayo, SD~Moffatt-Bruce. Alarm system management: evidence-based guidance
  encouraging direct measurement of informativeness to improve alarm response
  {\it {BMJ} Quality \& Safety. } 2015;24.

\end{thebibliography}
\bibliographystyle{ama}

\end{document}